\documentclass[number,5p]{elsarticle}
	\usepackage{amsmath}
	\usepackage{hyperref}

	\journal{Computers \& Fluids}
\newcommand\etal{\textit{et~al}\ }
\newcommand\figscale{0.7} 

\begin{document}

\begin{frontmatter}
	\title{Performance optimizations for scalable CFD applications on hybrid CPU+MIC heterogeneous computing system with millions of cores%
		}

	\author[nudt]{WANG Yong-Xian\corref{cor1}}
		\ead{yxwang@nudt.edu.cn}
	\author[nudt]{ZHANG Li-Lun}
		\ead{zll0434@sina.com}
	\author[nudt]{LIU Wei}
		\ead{liuweinudt@126.com}
	\author[nudt]{CHENG Xing-Hua}
		\ead{chengxinghua@nudt.edu.cn}
	\author[ttu]{Zhuang Yu}
		\ead{yu.zhuang@ttu.edu}
	\author[utsa,upg]{Anthony T. Chronopoulos}
		\ead{antony.tc@gmail.com}

	\cortext[cor1]{Corresponding author}

	\address[nudt]{National University of Defense Technology, Changsha, Hu'nan 410073, China}
	\address[ttu] {Texas Tech University, Lubbock, TX 79409, USA}
	\address[utsa]{University of Texas at San Antonio, San Antonio, TX 78249, USA}
	\address[upg]{Visiting Faculty, Dept Computer Engineering \& Informatics, University of Patras
26500 Rio, Greece}
	
	\begin{abstract}
	For computational fluid dynamics (CFD) applications with a large number of grid points/cells, parallel computing is a common efficient strategy to reduce the computational time. How to achieve the best performance in the modern supercomputer system, especially with heterogeneous computing resources such as hybrid CPU+GPU, or a CPU + Intel Xeon Phi (MIC) co-processors, is still a great challenge.
	An in-house parallel CFD code capable of simulating three dimensional structured grid applications is developed and tested in this study. Several methods of parallelization, performance optimization and code tuning both in the CPU-only homogeneous system and in the heterogeneous system are proposed based on identifying potential parallelism of applications, balancing the work load among all kinds of computing devices, tuning the multi-thread code toward better performance in intra-machine node with hundreds of CPU/MIC cores, and optimizing the communication among inter-nodes, inter-cores, and between CPUs and MICs. 
	Some benchmark cases from model and/or industrial CFD applications are tested on the Tianhe-1A and Tianhe-2 supercomputer to evaluate the performance. Among these CFD cases, the maximum number of grid cells reached 780 billion. The tuned solver successfully scales up to half of the entire Tianhe-2 supercomputer system with over 1.376 million of heterogeneous cores. The test results and performance analysis are discussed in detail.
	\end{abstract}

\end{frontmatter}


\section{Introduction}
\label{sec:1}

In recent years, with the advent of powerful computers and advanced numerical algorithms, computational fluid dynamics (CFD) has been increasingly applied to the aero\-space and aeronautical industries and CFD is reducing the dependencies on experimental testing and has emerged as one of the important design tools for these fields. 
Parallel computing on modern high performance computing (HPC) platform is also required to take full advantage of capability of computations and huge memories of these platforms when an increasing number of large-scale CFD applications are used nowadays to meet the needs of engineering design and scientific research.
It's well known that HPC systems are facing a rapid evolution driven by power consumption constraints, and as a result, multi-/many-core CPUs have turned into energy efficient system-on-chip architectures and HPC nodes integrating main processors and co-processors/accelerators have become popular in current supercomputer systems. The typical and widely-used co-processors and/or accelerators include GPGPUs (general purpose graphics processing unit), the first generation Intel (Knights Corner) MICs (Many Integrated Core), FPGA (field-programmable gate array), and so on. The fact that there are more than sixty systems with CPU+GPU and more than twenty systems with CPU+MIC in the latest Top500 supercomputer list release in November 2016 {indicates such a trend in} HPC platforms. 

Consequently, the CFD community faces an important challenge of how to keep up with the pace of rapid changes in the HPC systems. It could be hard, if not impossible, to port the high performing original efficient algorithm on the traditional homogeneous platform to the current new HPC systems with heterogeneous architectures seamlessly. The existing codes need to be re-designed and tuned to exploit the different levels of parallelism and complex memory hierarchies of new heterogeneous systems.

During the past years, researchers in the CFD fields have made a great efforts to implement efficiently CFD codes in the heterogeneous systems. Among many recent studies, researchers have paid more attention to CPU+GPU hybrid computing. 
Ref.~\cite{corrigan2011running} studied an explicit Runge-Kutta CFD solver for three-dimensional compressible Euler equations using a single NVIDIA Tesla GPU and got roughly 9.5x performance over a quad-core CPU. 
In \cite{griebel2010multi}, the authors implemented and optimized a two-phase solver for the Navier-Stokes equations using the Runge-Kutta time integration on a multi-GPU platform and achieved an impressive speedup of 69.6x on eight GPUs/CPUs.
Xu \etal proposed a MPI-OpenMP-CUDA parallelization scheme in \cite{xu2014balancing} to utilize both CPUs and GPUs for a complex, real-world CFD application using explicit Runge-Kutta solver on the Tianhe-1A supercomputer and achieved a speedup factor of about 1.3x when comparing one Tesla M2050 GPU with two Xeon X5670 CPUs and a parallel efficiency of above 60\% on 1024 Tianhe-1A nodes.

On the other hand, many studies aim to exploit the capability of CPU + MIC heterogenous systems for massively parallel computing of CFD applications. 
Ref.~\cite{saini2013early} gave a small-scale preliminary performance test for a benchmark code and two CFD applications on a 128-node CPU + MIC heterogeneous platform and evaluated the early performance results, where application-level testing was primarily limited to the case of using a single machine node. 
In the same year, authors in \cite{wangyx2013hpcc-alter} made an effort to port their CFD code on traditional HPC platform to the forthcoming new-setup Tianhe-2 supercomputer with CPU+MIC heterogeneous architecture, and the performance evaluations of massive CFD cases used up-to 3072 machine nodes of the fastest supercomputer in that year. 
Ref.~\cite{gorobets2014direct} reported simulations of running large-scale simulations of turbulent flows on massively-parallel accelerators including GPUs and Intel Xeon Phi coprocessors and found that the different GPUs considered substantially outperform Intel Xeon Phi accelerator for some basic OpenCL kernels of algorithm.
Ref.~\cite{vazquez2014xeon} implemented an unstructured meshes CFD benchmark code on Intel Xeon Phis by both explicit and implicit schemes and their results showed that a good scalability can be observed when using MPI programing technique. However, the openMP multi-threading and MPI hybrid case remains untested in their paper.
As a case study, Ref.~\cite{dengliang2015kepler} compared the performance of high-order weighted essentially non-oscillatory scheme CFD application on both K20c GPU and Xeon Phi 31SP MIC, and the result showed that when vector processing units are fully utilized the MIC can achieve equivalent performance to that of GPUs.
Ref.~\cite{smith2015performance} reported the performance and scalability of an unstructured mesh based CFD workflow on TACC Stampede supercomputer and NERSC Babbage MIC based system using up to 3840 cores for different configurations.

In this paper, we aim to study the porting and performance tuning techniques of a parallel CFD code to heterogeneous HPC platform. A set of parallel optimization methods considering the characteristics of both hardware architecture and the typical CFD applications are developed.The portability and device-oriented optimization are discussed in detail. The test result of the large-scale CFD simulations on Tianhe-2 supercomputer with CPUs + Xeon Phi co-processors hybrid architectures showed that a good performance and scalability can be achieved.

The rest of this paper is organized as follows: In Sect.~\ref{sec:2} the numerical methods, the CFD code and the heterogeneous HPC system are briefly introduced. The parallelization and performance optimization strategies of large-scale CFD applications running on heterogeneous system are discussed in detail in Sect.~\ref{sec:3}. Numerical simulations to evaluate the proposed methods and results and analysis are also reported in Sect.~\ref{sec:4}, and conclusion remarks are given in Sect.~\ref{sec:5}.
\section{Numerical Methods and High Performance Computing Platform}
\label{sec:2}

\subsection{Governing Equations}
\label{sec:2-1}
In this paper, the classical Navier-Stokes governing equation is used to model the three-dimensional viscous compressible unsteady flow. 
The governing equations of the differential form in the curvilinear coordinate system can be written as:
\begin{align}\label{eq:1}
    \frac{\partial \mathbf{Q}}{\partial t}
  +\frac{\partial (\mathbf{F} - \mathbf{F}_v)}{\partial \xi}
  +\frac{\partial (\mathbf{G} - \mathbf{G}_v)}{\partial \eta}
  +\frac{\partial (\mathbf{H} - \mathbf{H}_v)}{\partial \zeta}
  &= \mathbf{0},
\end{align}
where \(\mathbf{Q} = (\rho, \rho u, \rho v, \rho w, \rho E)^T\) denotes the conservative state (vector) variable, \(\mathbf{F}\), \(\mathbf{G}\) and \(\mathbf{H}\) are the inviscid (convective) flux variables, and \(\mathbf{F}_v\), \(\mathbf{G}_v\) and \(\mathbf{H}_v\) are the viscid flux variables in the \(\xi, \eta\) and \(\zeta\)  coordinate directions, respectively. Here, \(\rho\) is the density, \(u, v\) and \(w\) are the cartesian velocity components, \(E\) is the total energy. All these physics variables are non-dimensional in the equations, and for the three-dimensional flow field, they are vector variables with five components. The details of definition and expression of each flux variable can be found in \cite{deng1997compact}.

\subsection{Numerical Methods}
\label{sec:2-2}
For the numerical method, a high-order weighted compact nonlinear finite difference method (FDM) is used for the spatial discretization. Specifically, let us first consider the inviscid flux derivative  along the \(\xi\) direction. By using the fifth-order explicit weighted compact nonlinear scheme (WCNS-E-5) \cite{deng2011geometric}, its cell-centered FDM discretization can be expressed as 
\begin{align}
	\frac{\partial \mathbf{F}_i}{\partial \xi}
	= &\frac{75}{64h} ( \mathbf{F}_{i+1/2} - \mathbf{F}_{i-1/2} )
	- \frac{25}{384h} ( \mathbf{F}_{i+3/2} - \mathbf{F}_{i-3/2} ) \nonumber \\ 
	&+ \frac{3}{640h} ( \mathbf{F}_{i+5/2} - \mathbf{F}_{i-5/2} ),
\end{align}
where \(h\) is the grid size along \(\xi\) direction, and flux (vector) variable \(\mathbf{F}\) is computed by some kind of flux-splitting method by combining the responding left-hand and right-hand cell-edge flow variables. The discretization for other inviscid fluxes can be computed by a similar procedure. The fourth-order central differencing scheme is carefully chosen for the discretization of viscid flux derivatives to ensure all the derivatives have matching discretization errors. The reason why to design such a complex weighted stencil is to prevent numerical oscillation around discontinuities of flow field. 

When the flow field is composed of multiple grid blocks, in order to guarantee the requirement of high accuracy in numerical approach, the grid points shared by the adjacent blocks need be carefully dealt. As a result, the geometric conservation law proposed by Deng \etal \cite{deng2011geometric} and some severe conditions for the boundaries of the neighboring blocks must be met for complex configurations, see \cite{deng1997compact,deng2011geometric} for more details.

Once the spatial discretization is completed, we can get the semi-discretisation form of Eq.~(\ref{eq:1}) as follows:
\begin{align}\label{eq:3}
	\frac{ \partial \mathbf{Q} } {\partial t} = - \mathbf{R} (\mathbf{Q}).
\end{align}
In Eq.~(\ref{eq:3}), \( - \mathbf{R}(\mathbf{Q}) \) which denotes the discretized spatial items in Eq.~(\ref{eq:1}).
The discretization in time for the left hand side of Eq.~(\ref{eq:3}) will result into a multiple step advancing method.  In this paper, the third-order Runge-Kutta explicit method is used. Suppose superscript \((n)\) denote the \(n\)-th time step, and the time stepping can be expressed as:
\begin{align}
	\mathbf{Q}^{(n,1)} &= \mathbf{Q}^{(n)} + \lambda \cdot \mathbf{R}(\mathbf{Q}^{(n)}), \\
	\mathbf{Q}^{(n,2)} &= \frac{3}{4} \mathbf{Q}^{(n)} + \frac{1}{4} 
		\left[
			\mathbf{Q}^{(n,1)} + \lambda \cdot \mathbf{R}(\mathbf{Q}^{(n,1)})
		\right],\\
	\mathbf{Q}^{(n+1)} &= \frac{1}{3} \mathbf{Q}^{(n)} + \frac{2}{3} 
		\left[
			\mathbf{Q}^{(n,2)} + \lambda \cdot \mathbf{R}(\mathbf{Q}^{(n,2)})
		\right].
\end{align}

\subsection{Code implementation}
\label{sec:2-3}
\begin{figure}[htbp]
\centering
\includegraphics[scale=\figscale]{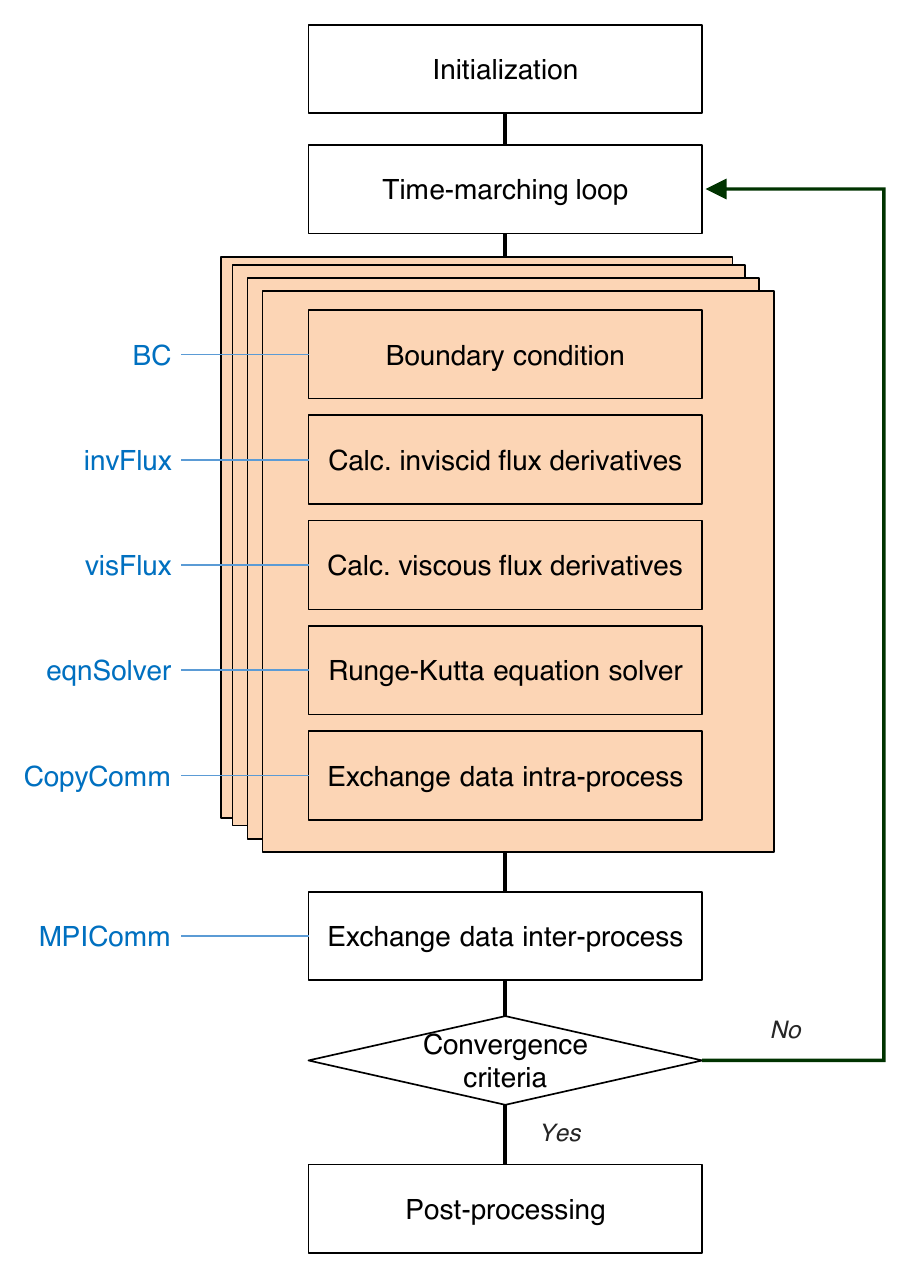}
\caption{Flowchart of CFD code}
\label{fig:1}
\end{figure}

An in-house CFD code capable of simulating three dimensional multi-block structured grid problem by using the  numerical method mentioned above is developed for the case study in this paper. The source code contains more than 15000 code lines of Fortran 90 programming language. In our previous work \citep{wang2013efficient}, the code is parallelized and tuned for the homogeneous HPC system composed of multi-/many-core CPUs. Shortly after that, we made a simple porting to CPUs + MICs heterogeneous system and conducted an early performance evaluation on the Tianhe-2 supercomputer \citep{wangyx2013hpcc-alter}. In this study, we will re-design and re-factor the code for the heterogenous many-core HPC platform to achieve a better overall performance. The flowchart of the CFD code is shown in Fig.~\ref{fig:1}. 

Similar to other CFD applications, our code consists of pre-processing, main and post-processing parts. At the pre-processing phase, the CFD code reads the mesh data and initializes the flow filed for the whole domain. Then the timing-consuming main part indicated by time-marching loop in the flowchart follows. In each time step iteration in this phase, the linear systems resulting from discretizing the governing equations in the domain are solved by numerical methods.  When the convergence criteria is met during the loop, the post-processing phase follows and finalizes the simulation.
The traditional parallelization method for the CFD program is mainly based on the technique of domain decomposition and it can be implemented on shared-memory and/or distributed-memory system employing respective programming models, such as OpenMP multi-threaded model and MPI multi-process model. In these implementations, the same solver is running for each domain, and some boundary data of neighboring domains need to be exchanged mutually during iterations.
\section{Parallelization and Optimization Methods}
\label{sec:3}

\subsection{Multi-level parallelization strategy considering both application features and computer architecture}
\label{sec:3-1}

In order to explore the potential capability of modern computer hardware and improve the scalability of applications, a typical CFD parallel simulation running on these systems is based on multi-level and/or multi-granular hybrid parallelizing strategy. 
Heterogeneous HPC system using more than one kind of processors or cores are becoming a trend. These systems gain performance or energy efficiency not just by adding the same type of processors, but by adding different accelerators and/or coprocessors, usually incorporating specialized processing capabilities to handle particular tasks.
For heterogeneous computer systems equipped with accelerators and/or coprocessors,  a common paradigm is ``CPU + accelerator and/or coprocessor'' heterogeneous programming model. For example, ``MPI + OpenMP + CUDA'' hybrid heterogeneous model is used when lots of GPU accelerators are employed, and  ``MPI + OpenMP + Offload'' hybrid heterogeneous model is used instead when lots of MIC coprocessors are employed. In both circumstances, a proper decision of how to divide the simulation tasks into lots of sub-tasks and map them to hardware units of the heterogeneous computer system should be carefully made in order to achieve good load balancing and high efficiency and parallel speedup.

\subsubsection{Principles of multiple level parallelization in typical CFD applications}
\label{sec:3-1-1}

Multi-level parallelization according to different granularities is the main means to improve performance of applications running on high-performance computers. In typical CFD applications, we try to exploit the parallelism at the following levels from the coarsest granularity to the finest granularity.

(1) Parallelism of multiple zones of flow field. 
At this level, the whole domain of flow filed to be simulated will be decomposed into many sub-domains and the same solver can run on each subdomain. Towards this goal, it is a common practice to partition the points/cells of discretized grid/mesh in the domain into many zones, as a result the parallelism of data processing among these zones can be obtained. 
It's worthy noting that for the purpose of workload balancing, it's far more than trivial to partition the grid/mesh and map the zones to hardware units in the modern  computer system which consists of a variety of heterogeneous  devices resulting different computing capabilities. 
However there are also some obstacles in the domain-decomposition. Among others, the convergence of implicit solver of linear system would be degenerated dramatically, which means more computational work is needed to make up for this, thus resulting more time-of-solution time.

(2) Parallelism of multiple tasks within each zone.
Considering the simulation process of each zone resulting from the domain-decomposition mentioned above, there are still a series of processing phases including updating convection flux variables and viscid flux variables in three coordinate directions, respectively. The fact that no data dependency exists among these six phases (i.e. 3 convection flux calculation phases plus 3 viscid flux calculation phases) implies the concurrency and the parallelism of multiple tasks.

(3) Parallelism of Single-Instruction-Multiple-data.
Even in each zone and each processing phase, the computing characteristic is highly similar among all grid points/cells, and lots of data distributed in different grid points/cells can be maneuvered by the same computer instruction. Besides the parallelization some architecture-oriented performance tuning techniques, such as data blocking, loop-based optimization, and so on, are also applied at this level.

\subsubsection{Parallelization of CFD applications in homogeneous systems}
\label{sec:3-1-2}

In typical homogeneous computing platforms, the multi-level parallel computing of traditional CFD applications is mainly implemented by the combination of MPI multi-process programming model, OpenMP multi-threaded programming model and single instruction multiple data (SIMD) vectorization techniques.

\begin{figure}[htbp]
\centering
\includegraphics[scale=\figscale]{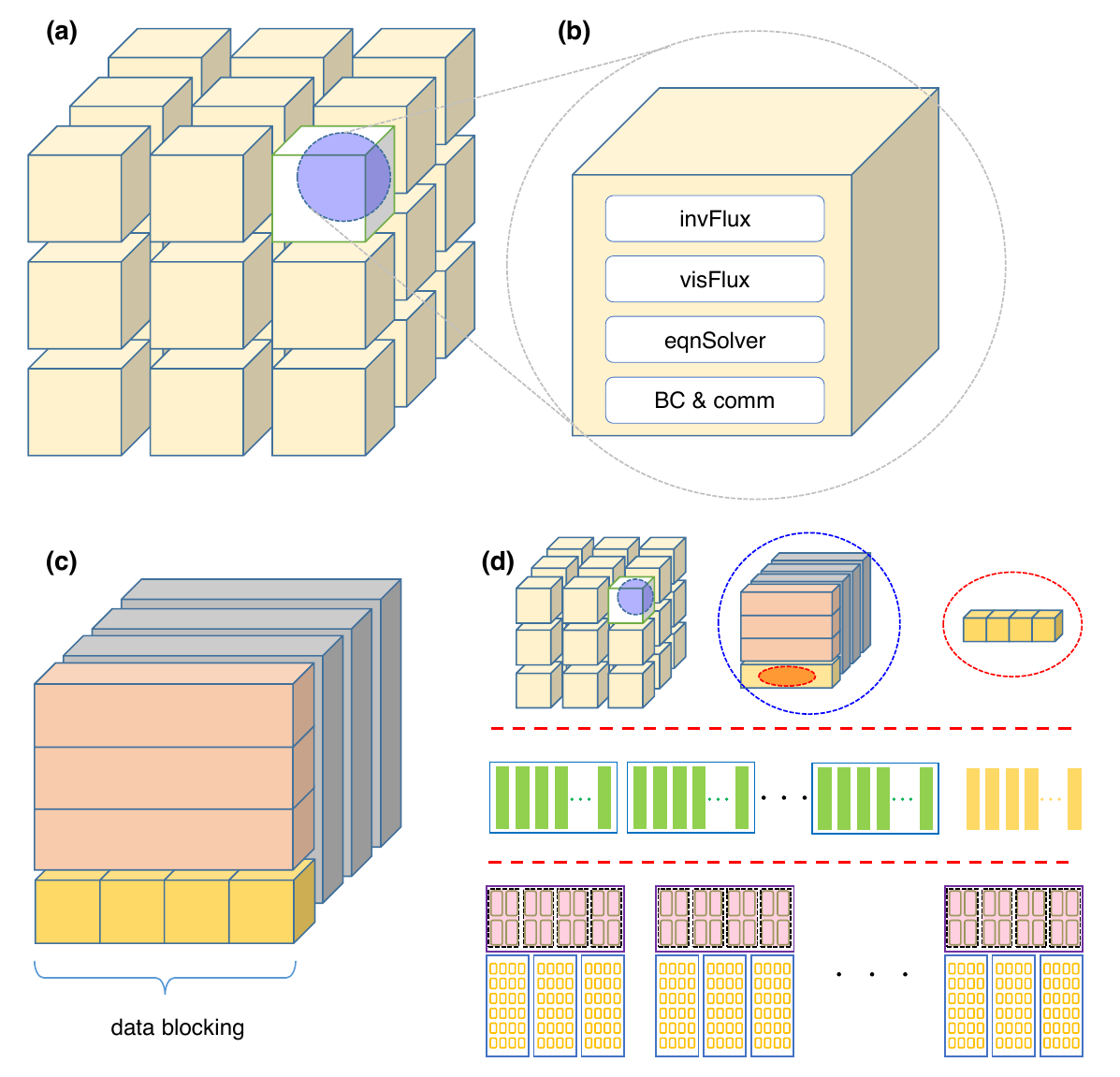}
\caption{Multi-granularity parallelization of typical CFD problems (a--c), and its specific multi-level mapping method (d)}
\label{fig:2}
\end{figure}

For the coarse-grain parallelism at the domain decomposition level, as shown in Fig.~\ref{fig:2}(a), the classical parallel computation uses a static mesh-partitioning strategy to obtain more mesh zones. For the case of commonly used structured grid, each mesh zone of flow field domain is actually a block of grid points/cells. In the following mapping phase, each mesh zone will be assigned to an executing unit, either a running process (in MPI case) or a running thread (in OpenMP multi-threading case), and these executing unit would further be attached to specific hardware units, namely CPUs, or cores.
In practice, a ``MPI + OpenMP'' hybrid method is applied in order to better balance the computation workload and avoid generating too small grid blocks (that usually leads too little workloads). This method does not need to over-divide the existing mesh blocks, and the workload can be fine-adjusted at thread-level by specifying a certain number of threads in proportion to the amount of points/cells in the block.

For the parallelism of intermediate granularity, as shown in Fig.~\ref{fig:2}(b), the multi-threading parallelization based on the tasks distributing principle using OpenMP programming model is applied. The most time-consuming computation for a single iteration of a typical CFD solver involves the calculation of the convection flux variables (the \verb|invFlux| procedure shown in Fig.~\ref{fig:1}), the viscous flux variables (the \verb|visFlux| procedure), large sparse linear system solver (the \verb|eqnSolver| procedure) and other procedures. The parallelization of this granularity can be further divided into two sub-levels: (1) at the upper sub-level, the computation in procedure \verb|invFlux| and \verb|visFlux| along three coordinate directions naturally forms six-task concurrent workloads, thus can be parallelized in the computing platform. (2) at the lower sub-level, considering the processing within each procedure, when discriminating between grid points/cells on the boundary of the zone and the inner ones of the zone, it can be easily found that by rearranging the code structure and adjusting the order of calculation, there is more potential parallelism to exploit based on some loop transformation techniques \cite{wang2013efficient}.

For the fine-grain parallelism within a single zone, as shown in Fig.~\ref{fig:2}(c), because the computation of each zone is usually assigned to a specific execution unit (process or thread), thus also mapped to a CPU core, the traditional SIMD (single instruction multiple data) parallelization can be applied to this level. Take an example, for the Tianhe-2 supercomputer system, as one of our test platforms, the CPU processor supports the 256-bit wide vector instruction, and the MIC coprocessor supports the 512-bit wide vector instruction, which provide great opportunity for us to exploit the parallelism at the instruction level based on the SIMD method. As seen in our previous experiences, the SIMD vectorization can usually bring us a 4X--8X performance improvement measured by the double-precision floating-point in a typical CFD application.

\subsubsection{Parallelization of CFD applications in heterogeneous systems}
\label{sec:3-1-3}

Heterogeneous system consists of more than one kind of processor or cores, which can results to huge differences of the computing capability, memory bandwidth and storage capability among different sub-systems. For the parallelization of CFD application in such heterogeneous systems, in addition to the implementation methods for the homogeneous platforms introduced in the previous section, more attention should be paid to the cooperation between main processors and accelerators/coprocessors. The strategy of load balancing, task scheduling, task distribution, and programming model in such circumstances should be carefully re-checked, or even re-designed. Now let's take the Tianhe-2 supercomputer system as an example. It consists of 16,000 machine nodes, and each node is equipped with both CPUs as the main processors and MICs as the coprocessors. The task distributions and collaborations between CPUs and MICs are implemented by so-called ``Offload programming model'' with the supported libraries provided by Intel company.
As illustrated in Fig.~\ref{fig:2}(d), task scheduling and load balancing can be divided into three layers. The computing tasks of multi-zone grid CFD application shown at the most top layer (namely ``application layer'') are firstly organized and assigned to multiple processes and/or threads (shown at ``executing layer''), thus mapped to specific hardware devices, say,  a core of either CPU or MIC at the ``hardware layer''. For simplicity and easy implementation, the task assigning,  scheduling and mapping between adjacent layers can be built in a static mode.
Considering the fact that each machine node of Tianhe-2 system contains two general-purpose CPUs and three MIC devices, in order to facilitate the balanced load between these two kind of computing devices, we designed the following configuration of parallelization:  multiple processes are distributed for each machine node via MPI programming model as well as many OpenMP threads are running within each process to take advantage of many-core features of both CPUs and coprocessors. Offloading computational tasks into three MIC devices is carried out by one of OpenMP threads, namely, main thread. Using this arrangement, each process can distribute some shares of its work to the corresponding MIC devices, and it is easier to utilize the capabilities of all MIC devices equipped on the machine nodes. The communication between CPUs and MIC devices is implemented by the pragmas or directive statements as well as support libraries provided by the vendor. In typical programming practice, the main thread of a CPU process will pick up some data to offload into MIC devices, and the MIC devices will return the resulting data back to the main thread running on CPU.
The load balancing between CPU and MIC is a big challenge in the applications, and a simple static strategy is applied in our implementation. We introduce a further intermediate layer in the ``application layer'' shown in Fig.~\ref{fig:2}(d) which organized the mesh zones into groups. As an example, five individual zones, say, $G_0$, $G_1$, $G_2$, $G_3$ and $G_4$, within a group of mesh zones, can further be divided into two sub-groups $\{\{G_0, G_1\}; \{G_2, G_3, G_4\}\}$ according their volumes and shapes, whereas $G_0$ and $G_1$ have almost equal volume and are assigned to two CPUs in the process respectively, and $G_2$, $G_3$ and $G_4$ have another near equal size and the computing tasks on them will be assigned to three MIC devices respectively. As an extra benefit of introducing the intermediate layer, regardless of the parallelization and optimization on CPU or on the MIC device, the strategy applied at finer-grain level, i.e. thread-level, SIMD level, etc, can be processed in a similar way.

\subsection{Implementations of parallelization and optimization for both homogeneous and heterogeneous platforms}
\label{sec:3-2}

\subsubsection{Load balancing method of parallel computing}
\label{sec:3-2-1}

The load balancing of parallel simulation of CFD applications mainly results from re-partitioning the existing mesh in the whole domain of flow field. The case of large-scale parallel CFD simulation imposed great requirements to the routine grid partitioning procedure. For example, when submitting such a simulation job to a specific HPC platform, the availability of computing resources, such as the number of available CPU nodes and/or cores, capability of memory system, etc, are varying and depending on the status in situ. Among others, a flexible solution is to partition the original mesh zone into many small-sized blocks followed by re-grouping these blocks into final groups of blocks. We developed a pre-processing software tool to fulfill this task, and the details of algorithm and implementation can be found in \cite{wang2013grid-alter}. In fact some complex re-partitioning techniques are introduced in the second processing for fine-adjusting the workload of computation and communication in the solution phase.

For the load balancing of OpenMP multi-threaded parallelization, there are two task scheduling strategies that are adopted in the CFD applications. The static task scheduling strategy based on distributing the workloads equally into threads is applied in the iteration parallel regions. However, for the concurrent execution of \verb|invFlux_X|, \verb|invFlux_Y|, \verb|invFlux_Z|, \verb|visFlux_X|, \verb|visFlux_Y|, \verb|visFlux_Z| procedures as described in Sect.~\ref{sec:3-1-2}, the dynamic task scheduling strategy is more appropriate due to the differences in the number of grid points/cells, the amount of computations among these procedures.

The load balancing between different computing devices, such as CPU and MIC, is another issue for the heterogeneous computing platform. The static task assigning method is applied to address this issue, and an adjustable ratio parameter, which denotes the ratio of amount of workload assigned on one MIC device to that on one CPU processor, is introduced to match the different computing platforms guided by the performance data measured in real applications. We will give the example and results for a series of test cases to discuss this issue in Sect.~\ref{sec:4-3}.

\subsubsection{Multi-threaded parallelization and optimization}
\label{sec:3-2-2}

\begin{figure}[htbp]
\centering
\includegraphics[scale=\figscale]{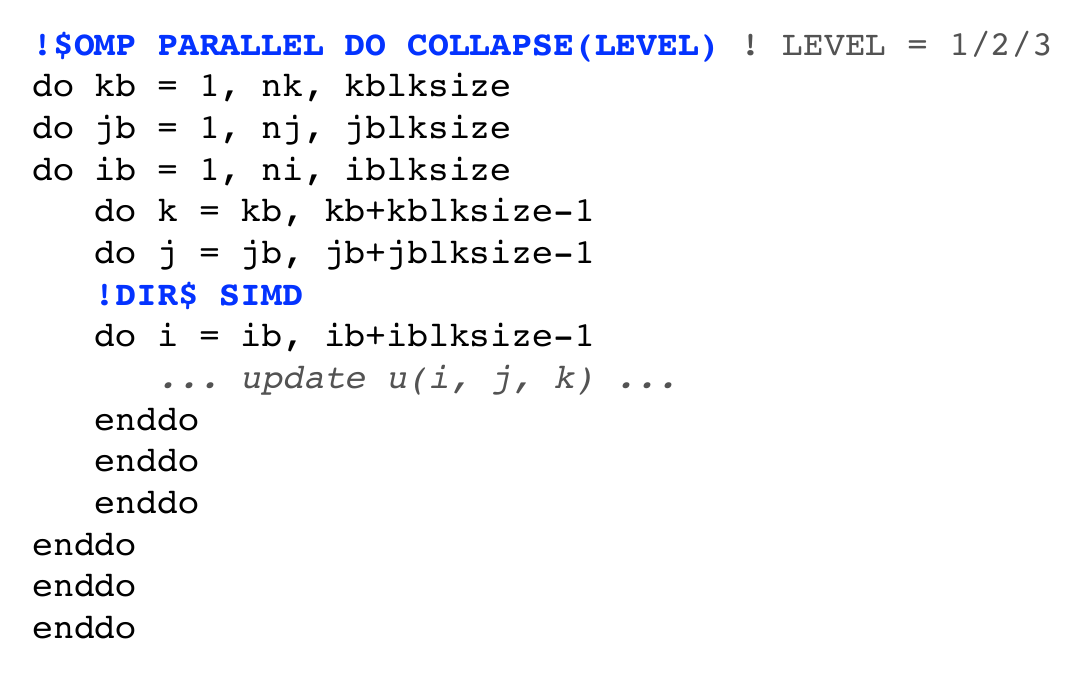}
\caption{Data blocking used in the parallelization and optimization }
\label{fig:3}
\end{figure}

It is well known that increasing number of processor cores are integrated in a single chip to gain the computing capability and make the balance between the power and the performance in modern computer platform. On these many-core computer system, multi-threaded programming model is widely used to effectively exploit the performance of all processor cores. Thus, the parallelization and optimization of multi-threading CFD simulation on many-core platform must be implemented. We emphasize some methods of parallelization and optimization as following:
(1) We must achieve a balance between the granularity and the degree of parallelism. In order to achieve this goal in typical CFD applications, the OpenMP multi-threading parallelization is applied for the computations of mesh points/cells within each single zone via splitting the iterative space of the zone in three coordinate directions, as illustrated in Fig.~\ref{fig:3}.
(2) We must reduce the overhead of thread creation and destruction. If possible, the OpenMP pragma/directive to create the multi-thread region should be placed at the outer loop of the nested loop to reduce the additional overhead caused by repeated dynamic creation and destruction of the thread. 
(3) We must reduce the amount of memory occupied by a single thread. As often done in general scientific and engineering computing, a lot of variables in the OpenMP multi-thread region will be declared as private variables to ensure the correctness of computations. As a result, the memory footprint of the application is too large to make the performance of memory access deteriorate. The issue is extremely severe in the MIC coprocessor device due to it has smaller storage capacity. To address this issue, we try to minimize the amount of memory allocated for private variable and let each thread only allocate, access and deallocate the necessary data  to obtain the maximum performance gains. 
(4) We must bind the threads to the processor cores in NUMA architecture. Ensuring static mapping and binding between each software thread and the hardware processor core in the computing platform with NUMA architecture can improve the performance of CFD application problems dramatically. This can be achieved by using a system call provided by the operating system, or the affinity clause provided by OpenMP library in CPU platform, and by setting some environment variables supported by vendors on the MIC platform.

\subsubsection{Multi-level optimization of communication }
 \label{sec:3-2-3}
 
In modern computer architectures, gains in processor performance, via increased clock speeds, have greatly outpaced improvements in memory performance, causing a growing discrepancy known as the ``memory gap'' or ``memory wall''. Therefore, as time has passed, high-performance applications have become more and more limited by memory bandwidth, leaving fast processors frequently idle as they wait for memory. As a result, to minimize the data communication, including data movement and data transfer, between different hardware units is an effective method of applications simulated on modern HPC platform.
On the other hand, it's well known that in traditional parallel CFD applications, there are lots of data communication, including data movement and data transfer, between different machine nodes, between different executing processes running on the same node, and between CPU core and MIC device, and so on. 
One of the keys to effectively decrease the cost of data communication is therefore to shorten the communication time as much as possible, or ``hide'' the communication cost by overlaying the communication with other CPU-consuming work.

In order to reduce the number of communication and the time cost per communication,  a reasonable ``task-to-hardware'' mapping is established. When assigning computation tasks of mesh grid to the MPI processes and scheduling the MPI processes on the machine nodes or processors in the running time, we prefer to distributing the adjacent blocks in spatial topology to the same process or the same machine node if possible, thus minimizing the occurrence number of communication across different nodes. If multiple communication between each pair of processes can be done concurrently, the messages will be packed and only one communication is needed. Furthermore, the non-blocking communication mode is used instead of blocking communication in conjunction with overlapping the communication and  the computation  to hide the cost of communication fully or in partially.

In large-scale CFD applications with multi-block mesh as well as high-order accuracy solver there is another issue with some neighboring mesh points. These so-called singular mesh points, located in common boundaries, are shared by three or more adjacent mesh blocks. When wider stencil schemes are used, as it is usually done in high-order accuracy CFD applications, more very small size messages are needed for communication, which thus brings additional overhead. An optimization method to address this issue is discussed in \cite{wangyx2013improved} by introducing non-blocking point-to-point communication.

Another main obstacle to prevent the performance improvement for CFD applications simulated in CPU + MIC heterogeneous platform by offloading programming model is that there is large amount of data need to be transferred forth and back between CPU and MIC coprocessor using the PCIe interface which has typically lower bandwidth comparing with the bandwidth of memory access. 
This can be optimized by two methods. 
Firstly we use asynchronous instead of synchronous mode to transfer the data between CPU and MIC coprocessor, as well as start the data transferring as soon as possible to partially or completely hide the data transfer overhead via overlapping the usage of CPU and the data transferring.
Secondly optimizing the communication between CPU and MIC coprocessor in order to reuse the data either transferred on the MIC device earlier or resulting from previous computations is applied as long as they can be reused in the following processing phase, such as in the next loop of an iterative solver. If only a part of data in an array would be updated in the MIC devices, we can use the compiler directive statement \verb|!Dir$ offload| supported in Intel Fortran compiler to offload only a slice of that array to the coprocessor. In other cases of transferring some induced variables which can be calculated from other primary variables, only those primary data are transferred to coprocessor along with offloading the computing procedures to calculate the induced data into coprocessor. This means that we recalculate the generated data on the coprocessor instead of transfer them between CPU and MIC device directly.

\subsubsection{The optimization of wide vector instruction usage}
\label{sec:3-2-4}

Both the CPUs or the MIC coprocessors of a modern computer system can support wide vector instruction set to utilize the SIMD (single instruction multiple data), or vectorization technique at instruction level. For example, in Tianhe-2 supercomputer, its CPU supports 256-bit wide vector instructions and its MIC coprocessor supports 512-bit wide vector instructions. Vectorization and its optimization is one of the key methods to improve the overall floating point computational performance. In our previous experience, it can bring us a 4x--16x speedup theoretically depending on the precision of floating point operations and the specific type of the processor used for the CFD simulation.
In practice of development, we use the compiler option, such as \verb|-vec| option in Intel Fortran compiler, in addition to the user-specified compiler directive statement to accomplish this goal. In fact, most modern compilers can make this kind of vectorization automatically, and what the user should do is checking the optimization report generated by the compiler, picking up the candidate code fragments not auto-vectorized, and adding proper advanced directives or declarations to vectorizing them manually providing to the correctness is ensured.

\subsubsection{CPU + MIC collaborative parallelization and optimization}
\label{sec:3-2-5}

\begin{figure}[htbp]
\centering
\includegraphics[scale=\figscale]{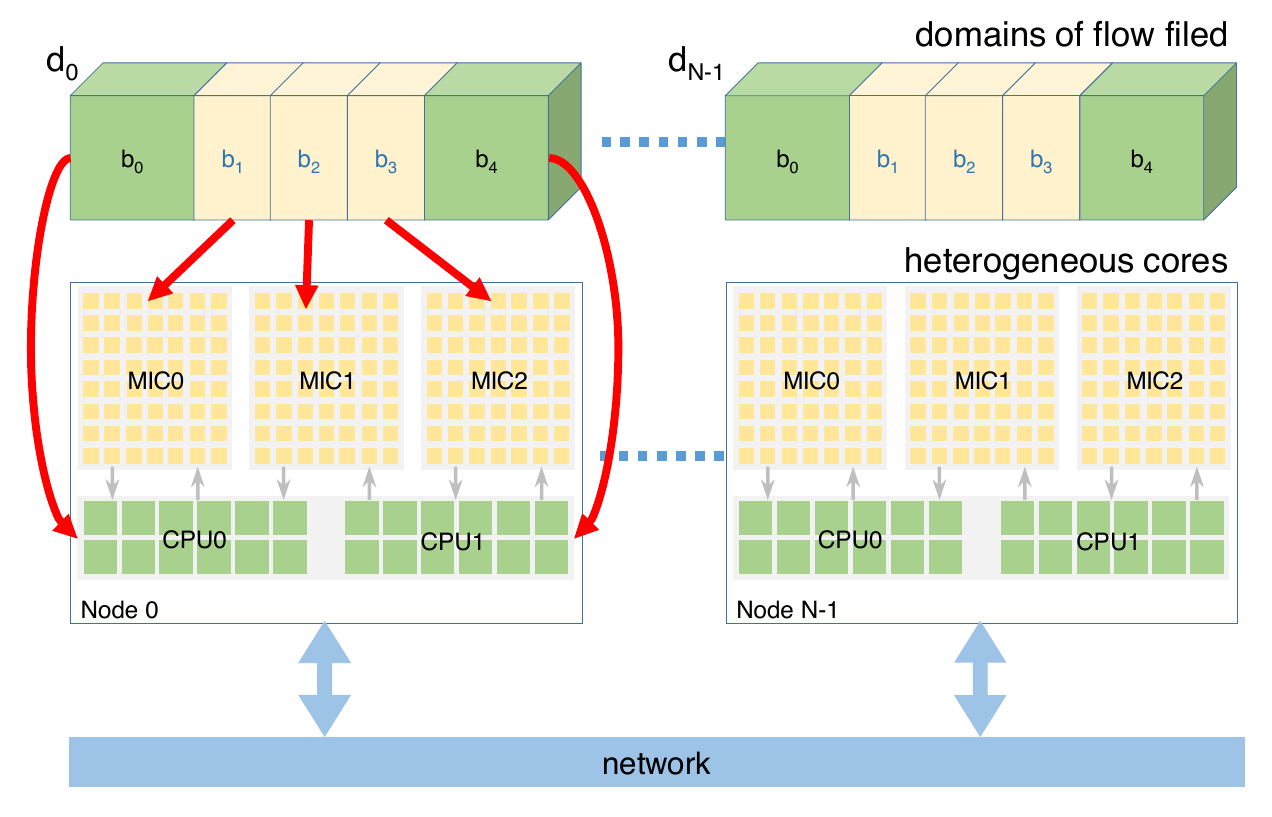}
\caption{Task assignments and mapping on CPU + MIC heterogeneous platforms}
\label{fig:4}
\end{figure}

There are significant differences in the capability of computation, storage capacity, network bandwidth and delay, etc. between different types of devices of heterogeneous platform, so we must carefully analyze, design and organize the appropriate order of computing, memory access and communication to achieve the collaborative parallel simulation for CFD applications in this heterogeneous platform. 
The collaborative parallelization is mainly applied within the single machine node of high-performance computer system. The load balancing between CPUs and MIC coprocessors is designed by assignment of amount of meshes and tasks as described in Sect.~\ref{sec:3-1-3}. All the grid blocks assigned to each process are divided into two types of groups in according to their volume and size, i.e. the total amount of mesh points/cells, whereas the first type of mesh groups, including two groups, are assigned to the CPUs of the machine node, and the second type of mesh groups, including the remaining three groups, are assigned to the MIC processors of the same machine node, as shown in Fig.~\ref{fig:4}. By adjusting the ratio of sizes of two types of grid blocks, the best value for load balancing can be obtained. On the CPU-side multiple threads can be created through the OpenMP programming model to utilize the capability of many-core processors. The main thread running on the CPU is responsible for interacting with the MIC devices, and the rest of the threads are responsible for the computing work distributed for the CPUs. A series of parallel optimizations of collaborative computing are applied based on this framework.

(1) Reducing the overhead of offloading tasks into MIC devices. 
Due to the limit bandwidth of communication between CPUs and MIC devices via PCIe interface, one of effective method to improve the performance is to increase the number and the volume of data transferred between these two types of devices. We can refactor the code lines and re-organize the small kernel to be executed on the MIC devices as a bigger kernel as possible as we can. By this method, all the tasks can be completed through a single offloading data instead of many offloading operations resulting to much more data movement overhead. Furthermore, in the usage of compiler directive statement \verb|!Dir$ offload| in the Fortran codes, adding \verb|alloc_if|, \verb|free_if| and other similar clauses as much as possible to avoid repeated allocating and deallocating space for the big arrays offloaded in the MIC device. Moreover the initialization and allocation of the variable should be arranged as early as possible, at least in the ``warm'' stage prior to the main iterations of computation.

\begin{figure}[htbp]
\centering
\includegraphics[scale=\figscale]{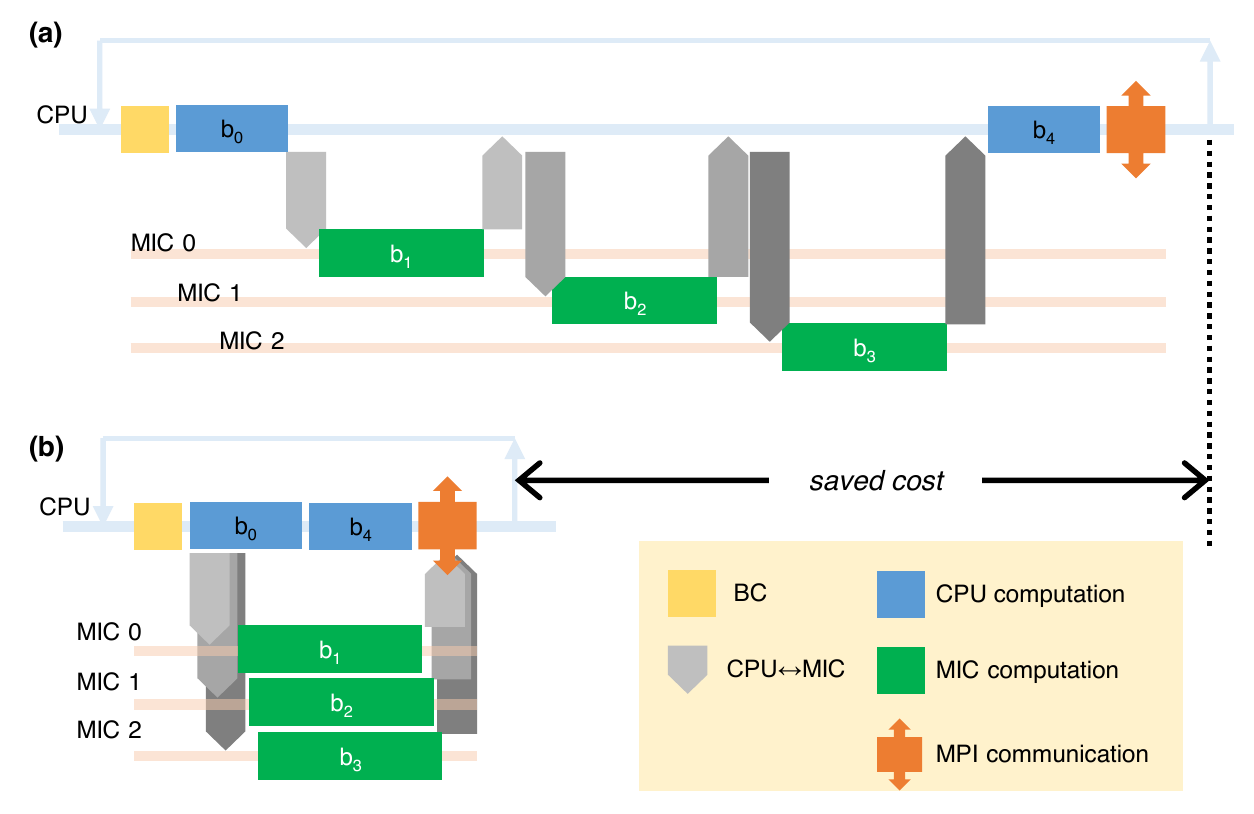}
\caption{Cooperative parallel optimization of CPU + MIC on a single node.
(a) no optimization.
(b) overlapping the computations and communication among CPUs and MICs.}
\label{fig:5}
\end{figure}

(2) Overlapping different levels of communication. 
In the flowchart shown in Fig.~\ref{fig:1}, once each \verb|eqnSolver| procedure is completed, it needs the data exchange and synchronization among all the blocks distributed on different processes and/or threads even within the same machine node. These communication can be further divided into two categories: inter-process MPI data communication and the data transferring between CPUs and MIC coprocessors. If the mesh partitioning is carefully designed based on the configurations of CPUs and MIC coprocessors, we can avoid any dependence between intra-process MPI communications and data transferring of two types of devices. This enables to overlap them via the asynchronous communication and further hide the communication overhead. For example, if one-dimensional mesh partitioning strategy is adopted as shown in Fig.~\ref{fig:4} and five neighboring blocks (or block groups), namely, $\{ b_i: i = 0, \ldots, 4\}$, are assigned to the same computing node. Let's also suppose that only left-most block $b_0$ and right-most bock $b_4$ are assigned to CPU processors, and the remaining three blocks $b_1, b_2, b_3$ are assigned to MIC coprocessors. It's apparent that only CPUs loading the data of both boundary block  $\{b_0, b_4\}$ require exchange their data with neighboring nodes by MPI communication, and the data exchange among $b_1$, $b_2$ and $b_3$ can be bounded in an intra-node way via offloading between CPUs and the MICs. As shown in Fig.~\ref{fig:5}, the total communication overhead can be effectively reduced due to this asynchronous overlapping technique.

(3) Parallelizing the processing of heterogeneous processors. 
There is essentially concurrency between two types of computing devices, say, CPU and MIC coprocessor, and among different devices of the same type, and that means if the computational tasks are loaded asynchronously among these heterogeneous devices, both CPUs and MICs, the parallelization of computing tasks and overall performance improvement can be gained. 

The communication overlapping and minimization of data movement at multiple levels and phases mentioned in this subsection can be illustrated in Fig.~\ref{fig:5}.

In addition to aforementioned parallelization and optimization, the traditional performance tuning methods for single processor, such as cache-oriented memory optimization, aligning the data, using asynchronous I/O operations, buffering, etc., are also suitable for large-scale parallel CFD applications. Another issue worthy noting is that in extremely large scale CFD simulations, which take a long time to complete, the mean time between failures (MTBF) of the high-performance computer system is decreasing dramatically. As the result, it is necessary to build a fault-tolerant mechanism both at system-level and at application-level. However this is beyond the discussion of this paper.
\section{Numerical Experiments and Discussions}
\label{sec:4}

\subsection{Platform and configurations of numerical experiments}
\label{sec:4-1}
In order to evaluate the performance of the various parallelization and optimization methods in the previous section, we designed and tested a series of numerical experiments including the early results tested on the Tianhe-1A supercomputer system, a typical many-core CPU homogeneous platform, and the latest results on Tianhe-2 supercomputer with CPU + MIC heterogeneous platform. The basic information of these two platforms is briefly summarized in Table~\ref{tab:1}. As shown in the table, Tianhe-2 system is composed of 16,000 computing nodes, and each node has heterogeneous architecture, including two Intel Xeon E5-2692 processors and three Intel Xeon Phi 31S1P coprocessors with Many Integrated Core (MIC) architecture. The CFD application used for the large-scale parallel test is an in-house code aiming to simulating the problems with multi-block structured grid, which is developed using Fortran 90 programming language as well as Intel fortran
compiler of version 13. The \verb|-O3| compiling option is applied on CPU unless otherwise noted in all test cases, and the \verb|-O3 -xAVX| compiling options are applied for the test on CPU + MIC heterogeneous platform in order to generate correct vectorization instruction for the coprocessor.
There are four configurations of test cases in the following discussion. 
(1) ``DeltaWing'' case is for simulation of flow field around a  delta wing, and has 44 grid blocks and a total of 2.4 million of grid cells.
(2) ``NACA0012'' case is for simulation of flow field around the NACA0012 wing, which has a single block of grid and a total of 10 million of grid cells.
(3) In the ``DLR-F6'' case, the amount of grid cells is 17 millions.
(4) In the ``CompCorner'' case (Fig.~\ref{fig:13}), the flow field for simulating the compressible corner problem are computed \cite{libangming2012theo} for a variety of problem sizes measured by the amount of grid cells. For this purpose, a special three-dimensional grid generating software, which can vary the total amount of grid cells, the number and the connecting structure of the grid blocks, and so on, is developed to generate different configurations for the test. 

\begin{table*}
\caption{The configurations of Tianhe-1A and Tianhe-2 platforms}
\label{tab:1}
\footnotesize
\centering\begin{tabular}{lll}
\hline
& Tianhe-1A & Tianhe-2 \\
\hline
CPU &
	Intel Xeon X5670 (6 cores per CPU) &
		Intel Xeon E5-2692 (12 cores per CPU) \\
Frequency of CPU &
	2.93 GHz &
		2.2 GHz \\
Configuration per node &
	2 CPUs &
		2 CPUs + 3 MICs \\
Memory capacity per node &
	48 GB &
		64 GB for CPUs + 24 GB for MICs \\
Coprocessors used &
	not used in this paper &
		Intel Xeon Phi (MIC) 31S1P (57 cores per MIC)\\
\hline
\end{tabular}
\end{table*}

All performance results reported in this section are the best one taken from five independent tests, and the timing results is only for the phase of main iterations in the flowchart shown in Fig.~\ref{fig:1}. We also limit the number of the main iteration to no more than 50 for the purpose of performance comparison.

\subsection{Performance results on homogeneous platforms}
\label{sec:4-2}
The Tianhe-1A supercomputer system, as a typical homogeneous HPC platform, is used in this subsection for the performance evaluations of following tests.

\begin{figure}[htbp]
\centering
\includegraphics[scale=\figscale]{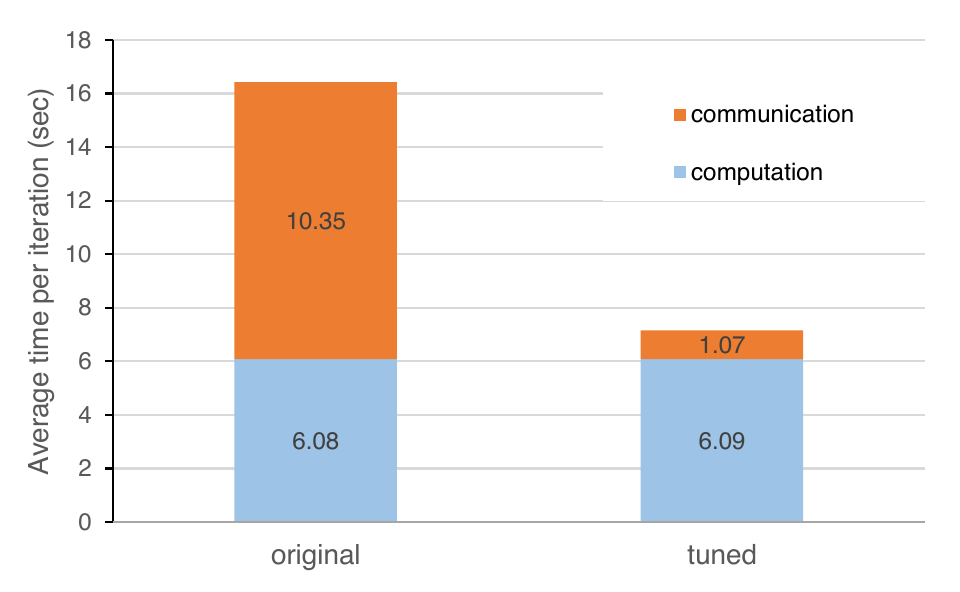}
\caption{Comparison of ratio of computation to communication between original and tuned version}
\label{fig:6}
\end{figure}

We firstly design a group of numerical experiments to test the methods of load balancing and communication optimization proposed in Sect.~\ref{sec:3-1}, and the ``NACA0012'' case with 10 million of mesh cells is employed on Tianhe 1A supercomputer system. For this purpose, the test uses 64 symmetric MPI processes, one for each machine node, and we evaluate the average wall time of both computation and communication per iteration for the two version of applications, namely, without optimization and with optimization, respectively. The performance results are shown in Fig.~\ref{fig:6}.
In the CFD simulation of original version (without optimization), the time of communication among MPI processes is about 63\% of total time per iteration. In the tuned version, as described in Sect.~\ref{sec:3-1}, we eliminate redundant global communication operations, maximize the use of non-blocking communication by the help of refactoring the codes and overlapping communication and computation as much as possible. These optimizations significantly reduce the total overhead of communication, resulting to nearly 10 times increasing of the ratio of computation to communication.

For the multi-threaded parallelization and its optimization discussed in Sect.~\ref{sec:3-2}, among other methods, the affinity (or the binding) of OpenMP threads to CPU cores on the impact of parallel CFD application performance is the focus in our tests. The ``DeltaWing'' case are employed on the Tianhe-1A platform for this purpose. 
The results (not shown here) indicate that using the binding strategy of the thread to the CPU core can significantly improve the performance of parallel simulation. The more the number of threads used per process, the greater the performance gain. 
Additional numerical experiments for evaluating the impact of different methods in thread-level parallel optimization proposed in Sect.~\ref{sec:3-2}, such as data blocking, reducing memory footprint, etc, are also conducted (the results is not shown here). Although the results indicate some minor performance improvement benefit from these optimization, the threads affinity strategy shows better gains instead.

\begin{figure}[htbp]
\centering
\includegraphics[scale=\figscale]{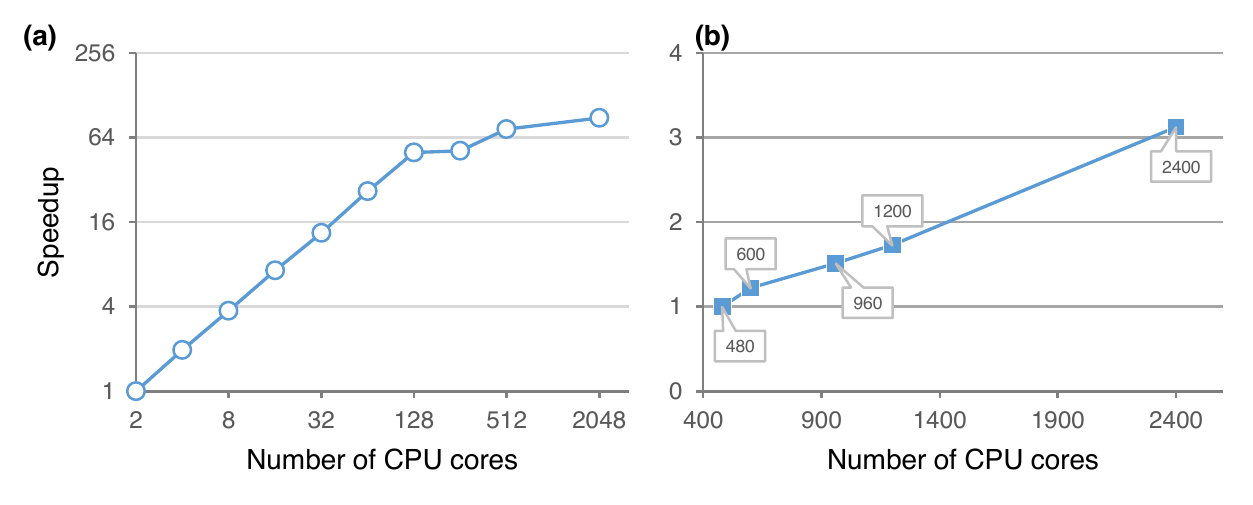}
\caption{Parallel speedup curves for two examples:
(a) DLR-F6 case with 17 million mesh cells.
(b) CompCorner case with 800 million mesh cells.}
\label{fig:8}
\end{figure}

To assess the overall effect of optimization method proposed on Sect.~\ref{sec:3-2} for the homogeneous system, the performance of two test cases with different runtime configurations on the Tianhe-1A system are reported in Fig.~\ref{fig:8}, where the horizontal axis is the number of CPU cores used in the test, and the vertical axis is the relative speedup to the baseline. In Fig.~\ref{fig:8}(a), DLR-F6 case with 17 million of grid cells is tested and the baseline is the performance when using two CPU cores. As a larger scale case, the CompCorner case with 800 million of grid cells is used in Fig.~\ref{fig:8}(b) and the performance in the case of using 480 CPU cores is taken as the baseline.
In both tests, two MPI ranks, six threads for each, are created for each machine nodes to maximizing the CPU cores of each machine node.
As can be seen from Fig.~\ref{fig:8}(a), for a medium-sized CFD application like the DLR-F6 case, the parallel speedup is basically linear growing when the number of CPU cores is no more than 256. However, further increasing of the number of CPU cores result to the decreasing of parallel speedup significantly due to the lower ratio of computation to communication for each CPU core. More specifically, there are two reasons. Firstly, increasing the number of processes increases significantly the number of singular points shared by neighboring intra-node MPI processes, which leads to much more overhead of MPI communication resulting from a large number of tiny-sized messages to be transferred among machine nodes \cite{wangyx2013improved}. Secondly, for a fixed-size CFD problem, using more CPU cores means decreasing the size of sub-problem running in each CPU core (in typical situation, the number of grid cells distributed in each CPU core is less than 10,000) as well as the size of sub-problem of each thread, which further leads to the degradation of parallel performance. 
In Fig.~\ref{fig:8}(b), the test case has a larger size of 800 million mesh cells, running on the configurations of 480, 600, 960, 1200, and 2400 CPU cores, respectively. Since the problem size in each grid block is large enough (more than 10 million of the grid cells), the ratio of computation to communication is relatively high, and the relative parallel speedup has a near linear relationship with respective to the number of CPU cores used in the simulations.

\begin{figure}[htbp]
\centering
\includegraphics[scale=\figscale]{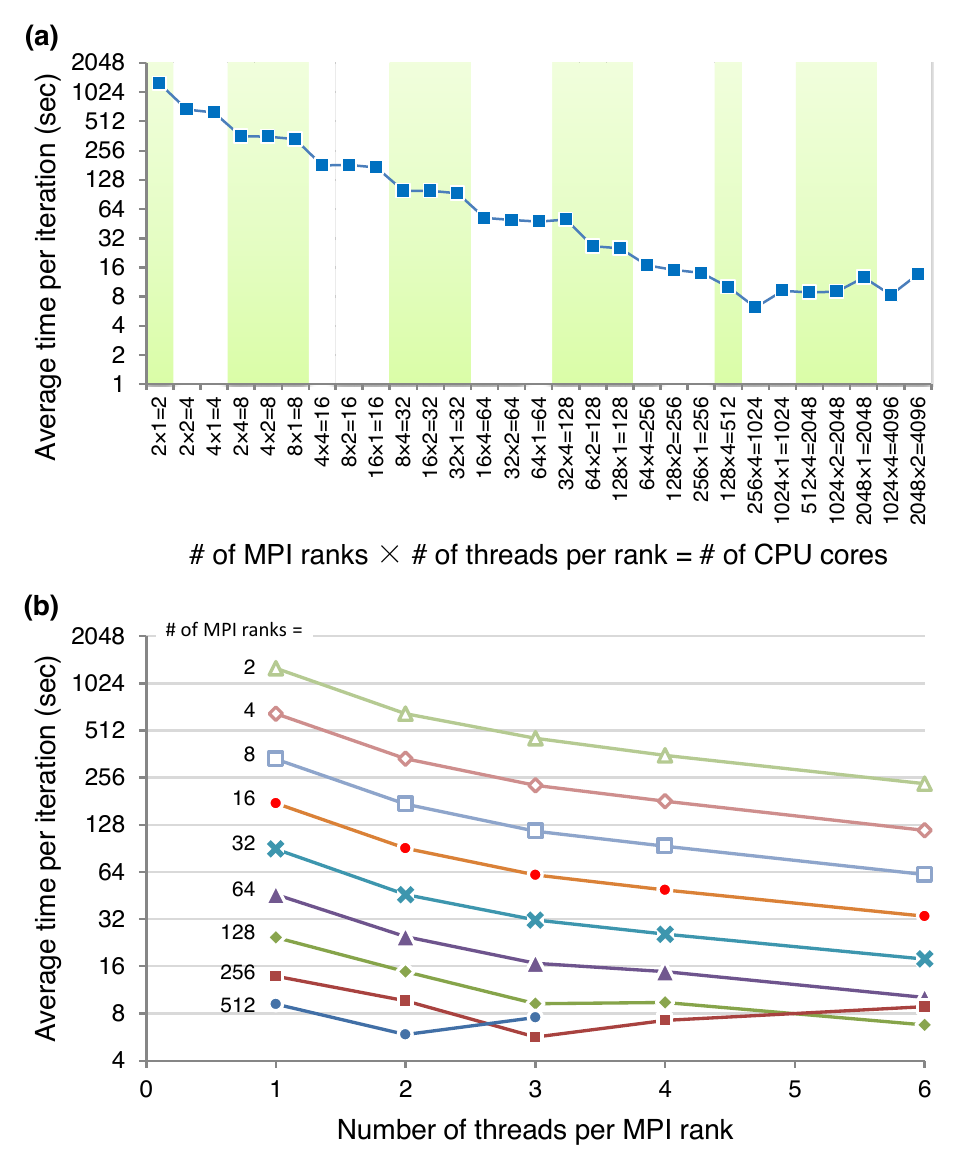}
\caption{Performance results in different combinations of processes and threads.
(a) Running time v.s. configurations of MPI ranks and OpenMP threads.
(b) Running time v.s. number of threads per MPI process.}
\label{fig:9}
\end{figure}

In order to compare the parallel performances of CFD simulation under different configurations with a variety of processes and threads combinations, we conducted a comprehensive tests for the medium-sized DLR-F6 case (17 million of grid cells), as shown in Fig.~\ref{fig:9}. 
Fig.~\ref{fig:9}(a) shows the relationship between running time per iteration and the number of CPU cores used in different configurations of MPI ranks and OpenMP threads, where we omit the results when the number of threads per process exceed 4. In each specific combination, the total number of CPU cores is ensured to be the same as the number of total number of threads. Some facts can be seen from the results. Firstly, when the number of CPU cores is less than 256, where a linear parallel speedup is observed, there is no significant performance variance among different combinations of processes and threads as long as the total number of cores is fixed. Secondly, when we limit the maximum number of threads per process to no more than 3, the parallel efficiency will be high as the number of cores increases, and then it declines  after the number of cores reaches 256. As a contrast, when limiting the maximum number of threads per process to no more than 4, the critical point of parallel efficiency will be extended to 1024 CPU cores. In both cases the number of MPI processes are limited to 256, and then the parallel efficiency begins to decline. This also shows that the cost of communication among MPI processes becomes the main obstacle of scaling the parallel simulation further.

In Fig.~\ref{fig:9}(b) we rearrange the performance results as the running time v.s. the number of threads per MPI process. It shows that only by increasing the number of threads to improve the parallel performance still has an upper limit. For example, in the case of using 256 MPI processes, the best performance can be achieved when 3 threads per process used. However, in the case of 512 MPI processes, the performance has no any further improvement when more than 2 threads per process used. The main reason is still the existing lower ratio of computation to communication in each MPI process when using more and more threads. In fact, for the case of 512 processes, only about 30,000 grid cells are distributed to each each process, and it is indeed a light load for the powerful CPU. If more threads are used in this case, the additional overhead it brings about is higher than the performance gain from computation acceleration, thus the overall performance declines.

\subsection{Performance results on heterogeneous platforms}
\label{sec:4-3}
We use the collaborative parallelization methods proposed in Sect.~\ref{sec:3-2} for the testing on the CPU + MIC heterogeneous platform. That is, only one MPI process is running on each machine node, as well as OpenMP multi-threaded running in each MPI process for finer parallelization, among those threads, one thread, named main thread, is responsible for offloading and collecting the sub-task to/from the three MIC devices within the machine node. 
During the phase of task assignment, each process reads in five grid blocks, two of which take the same size and are assigned to two CPUs, and the other three blocks with another size are assigned to the three MICs within the same node.

\begin{figure}[htbp]
\centering
\includegraphics[scale=\figscale]{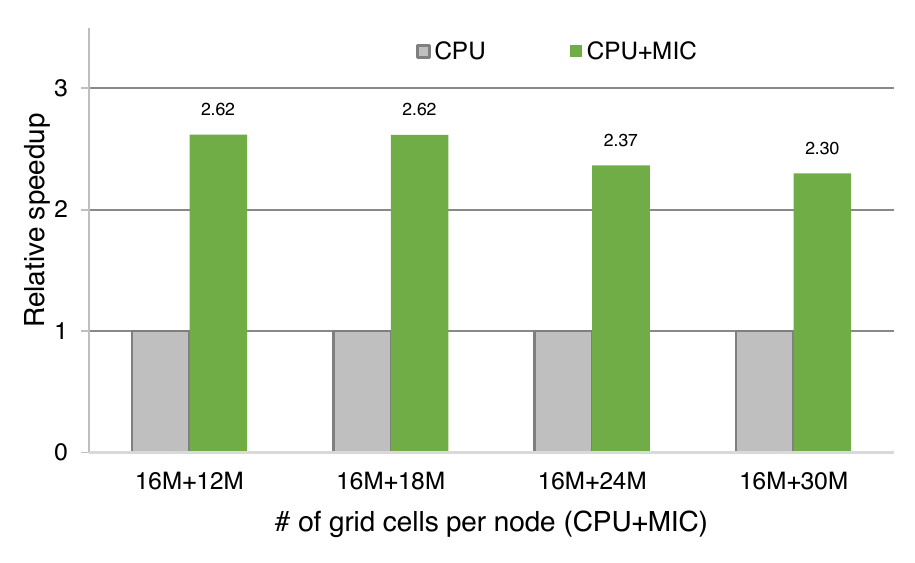}
\caption{Speedup relative to CPU-only test for four configurations.}
\label{fig:10}
\end{figure}

\begin{figure}[htbp]
\centering
\includegraphics[scale=0.5]{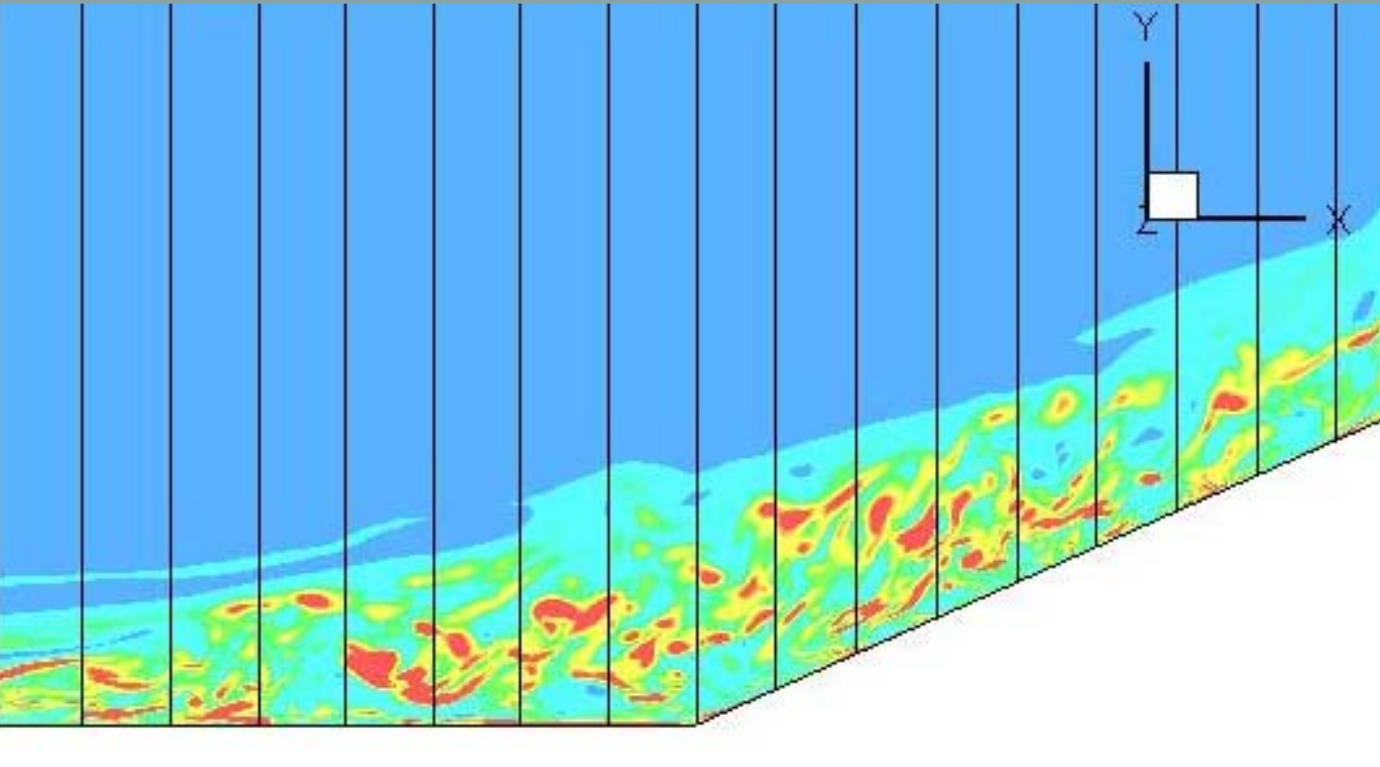}
\caption{Flow field result near the corner area in the simulation of test case ``CompCorner''}
\label{fig:13}
\end{figure}

To find the best load balancing between two types of devices, that is, CPUs and MICs, we firstly fixed the grid size as 8 millions (8M) for each CPU, thus 1600M for both CPUs in a machine node, and varied the grid size from 4M, 6M, 8M to 10M for each MIC coprocessor, respectively. As a contrast, each test of aforementioned heterogeneous computing will be accompanied by another test with the same total grid size but using only the CPU devices in the same machine node.
The acceleration results of all these four pairs of tests using ``CompCorner'' case on 16 Tianhe-2 nodes are reported in Fig.~\ref{fig:10}, and the flow field near the corner area are shown in Fig.~\ref{fig:13}. It shows that when the grid size for each MIC device is about 4M--6M, the optimal acceleration about 2.62X can be achieved for heterogeneous computation than the CPU-only running.

\begin{figure}[htbp]
\centering
\includegraphics[scale=\figscale]{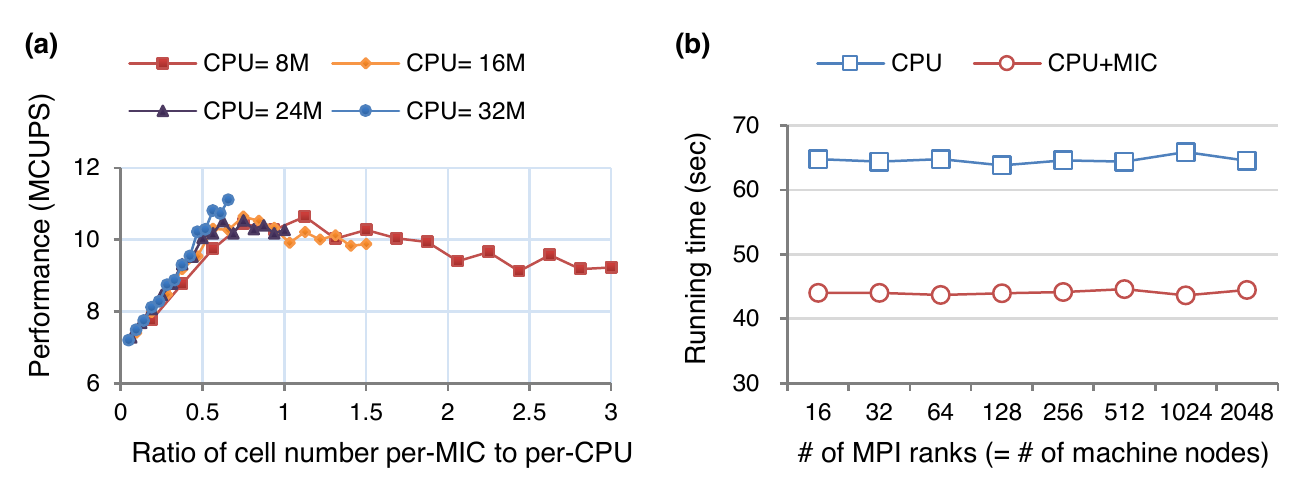}
\caption{Balancing the load between CPUs and MIC coprocessors.
(a) Performance results under different values of load ratio parameter.
(b) Scaling up to 2048 machine nodes with load ratio 0.6.}
\label{fig:11}
\end{figure}

To further study the optimal ratio of MIC workload to CPU load in heterogeneous computing, in the following a series of tests we also vary the grid size for each CPU from 8M, 16M, 24M to 32M. Fig.~\ref{fig:11}(a) shows the results of performance measure of million cell updates per second (MCUPS) when the ratio of the grid size per CPU to the grid size per MIC varies. It seems that when the ratio value is between 0.6 and 0.8, the highest performance can be observed in each group of tests, which is also nearly consistent with the results shown in Fig.~\ref{fig:10}.
In Fig.~\ref{fig:11}(b) we fix 16M grid cells for each CPU and 9.6M grid cells for each MIC (thus the load ratio is 0.6) and scale the problem size up to 2048 Tianhe-2 machine nodes. The observed good weak scalability confirms the effectiveness of methods of performance optimizations proposed in Sect.~\ref{sec:3-2}.

\begin{figure}[htbp]
\centering
\includegraphics[scale=\figscale]{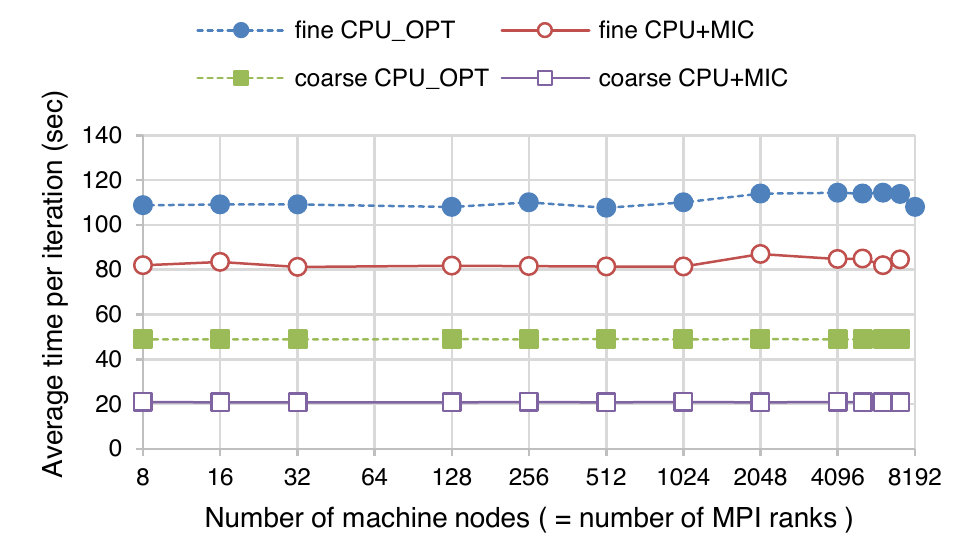}
\caption{Weak scalability for two platforms and two testing cases}
\label{fig:12}
\end{figure}

As a larger-scale test of our parallelization and optimization for both CPU-only homogeneous platform and CPU + MIC heterogeneous platform, the ``CompCorner'' cases with two types of configurations are used again on the Tianhe-2 supercomputer system. In the first type of configuration, shown as ``coarse'' in Fig.~\ref{fig:12}, each machine node process 40 million (40M) of grid cells, and 95.2M of grid cells in the ``fine'' type of configuration. In the CPU-only homogeneous tests, the number of machine nodes is up to 8192 nodes with total 196,600 CPU cores, and the largest problem size in ``fine'' configuration reaches 780 billion of grid cells. In the CPU + MIC heterogeneous tests, the maximum number of machine nodes is 7168 with 1.376 million of CPU + MIC processor/coprocessor cores, and the largest case has 680 billion of grid cells. The weak scalability results in Fig.~\ref{fig:12} shows the running time changing with the problem size, or equivalently, the number of machine nodes. From the result we can find, whatever the platform type, either CPU-only homogeneous system or CPU + MIC heterogeneous system, and the load per node, either as in the ``coarse'' configuration or as in the ``fine'' configuration, there is basically no change in running time when the problem size is increasing in proportion to the computing resources, thus showing a very good weak scalability.

\section{Conclusions}
\label{sec:5}

In this paper, the efficient parallelization and optimization method for large-scale CFD flow field simulation on modern homogenous and/or heterogeneous computer platforms is studied focusing on the techniques of identifying potential parallelism of applications, balancing the work load among all kinds of computing devices, tuning the multi-thread code toward better performance in intra-machine node with hundreds of CPU/MIC cores, and optimizing the communication among inter-nodes, inter-cores, and between CPUs and MICs. A series of numerical experiments are tested on the Tianhe-1A and Tianhe-2 supercomputer to evaluate the performance.
Among these CFD cases, the maximum number of grid cells reaches 780 billion. The tuned solver successfully scales to the half system of the Tianhe-2 supercomputer with over 1.376 million of heterogeneous cores, and the results show the effectiveness of our propose methods.
\section*{Acknowledgments}

We would like to thank NSCC-Guangzhou for providing access to the Tianhe-2 supercomputer as well as their technical guidance. This work was funded by the National Natural Science Foundation of China (NSFC) under grant no. 61379056.
\section*{References}
\bibliography{parcfd2017.bib}

\end{document}